\documentclass[twocolumn,showpacs,preprintnumbers,amsmath,amssymb]{revtex4}
\usepackage{graphicx}
\begin{document}
\title{Calibration of a solid state nuclear track detector (SSNTD) with high detection threshold to search for rare events in cosmic rays}
\author{S. Dey$^1$, D. Gupta$^{1,2}$,  A. Maulik$^1$\footnote{Corresponding author. Fax: +91 3325693127.\\E-mail address:atanu@bosemain.boseinst.ac.in(Atanu Maulik).}, Sibaji Raha$^{1,2}$, Swapan K. Saha$^{1,2}$}
\affiliation {$^1$Centre for Astroparticle Physics and Space Science,~Bose Institute, Kolkata 700 091, India}
\affiliation{$^2$Department of Physics, Bose Institute, Kolkata 700 009, India}
\author{D. Syam}
\affiliation{Department of Physics, Barasat Government College, Kolkata 700 124, India}
\author{J. Pakarinen, D. Voulot, F. Wenander}
\affiliation{CERN, CH-1211 Geneve 23, Switzerland}

\date{\today}
\begin{abstract}
We have investigated a commercially available polymer for its  suitability as a solid state nuclear track detector (SSNTD). We identified that polymer to be polyethylene terephthalate (PET) and found that it has a higher detection threshold compared to many other widely used SSNTDs which makes this detector particularly suitable for rare event search in cosmic rays as it eliminates the dominant low Z background. Systematic studies were carried out to determine its charge response which is essential before any new material can be used as an SSNTD. In this paper we describe the charge response of PET to $^{129}$Xe, $^{78}$Kr and $^{49}$Ti ions from the REX-ISOLDE facility at CERN, present the calibration curve for PET and characterize it as a nuclear track detector.  
\end{abstract}
\maketitle

\section{Introduction}

Solid State Nuclear Track Detectors (SSNTDs) have been used for many years for detecting charged particles in a wide range of disciplines~\cite{FL75,DU87}.
Charged particles on their passage through SSNTDs leave behind narrow damage trails. On being treated with suitable chemical reagents (etchants), material along the damage trails are etched out at a much faster rate (called track etch rate $V_T$) compared to the rate of etching of the undamaged bulk material (called bulk etch rate $V_B$). The resulting etch pits can be approximated by geometrical cones with the damage trails as axes and can be viewed under an optical microscope. The study of etch pit geometry and determination of the range of particles in the detector can reveal the identity of the particles forming the tracks and also their energies.

Their simplicity, ruggedness and existence of thresholds for registration make SSNTDs particularly suitable when it comes to the search for rare exotic particles like strangelets \cite{klingenberg,SA04} in cosmic rays. This is because such an experiment requires the deployment of large area detector arrays at remote terrains. We plan to deploy such SSNTD arrays at very high mountain altitudes, where theoretical studies~\cite{BA00} point to the presence of a measurable strangelet flux.

We have investigated a commercially available polymer (Century de'Smart, India) and found that it could be used as an SSNTD \cite {basu2005}. In an earlier work~\cite {RadMeas1} we identified the polymer material to be polyethylene terephthalate (PET) by chemical analysis and FTIR spectroscopy with chemical formula (C$_{10}$H$_{8}$O$_{4}$)$_n$. We have also shown that it has a much higher detection threshold compared to other widely used SSNTDs like CR-39, Lexan etc. We carried out systematic studies on PET to determine its bulk etch rate, ideal etching condition and also its charge response characteristics to $^{16}$O, $^{32}$S, $^{56}$Fe, $^{238}$U ions ~\cite{maulik, nimb}. Higher detection threshold makes PET an ideal choice as detector material for rare event search in cosmic ray research as it effectively suppresses the dominant low Z background. Another advantage of PET is its low cost compared to other commercially available SSNTDs. 

Before a new detector material can be employed, it needs to be properly characterized and calibrated. In this paper we report the results of charge response studies using $^{129}$Xe, $^{78}$Kr and $^{49}$Ti beams from the REX-ISOLDE~\cite{voulot} facility at CERN and establish the calibration curve for PET incorporating the new data. Present work along with our previous investigations firmly establishes PET as an SSNTD particularly suited to be used for rare event search in cosmic rays.

\section{Study of charge response of PET}

In order to study the charge response of PET, detector films were exposed to different ion beams from particle accelerators. The charge response or reduced etch rate ($V_T/V_B$) of SSNTDs depends on the specific energy loss ($dE/dx$) of the incident charged particle and hence is related to $(Z/\beta)$ by the Bethe equation~\cite{knoll}, where Z is the atomic number of the incident particle and $\beta$=v/c, gives a measure of its velocity. The primary consideration while carrying out the exposures was to make sure that the incident ions on PET have energies higher than the Bragg peak energy for that particular ion in PET. It is because at energies lower than that of the Bragg peak, the Bethe equation starts to break down due to charge neutralization. Beam currents and exposure durations were chosen so that the number of ions impinging on the detector lies in the range $\sim 10^4-10^5~/cm^2$. This is to prevent overlapping of tracks and detector burnout and also to optimize the data analysis process.

\vspace{0.3 cm}
\centerline{\bf{EXPERIMENT}}
\vspace{0.3 cm}

\begin{figure}[h]
\includegraphics[width = 1.0\hsize,clip]{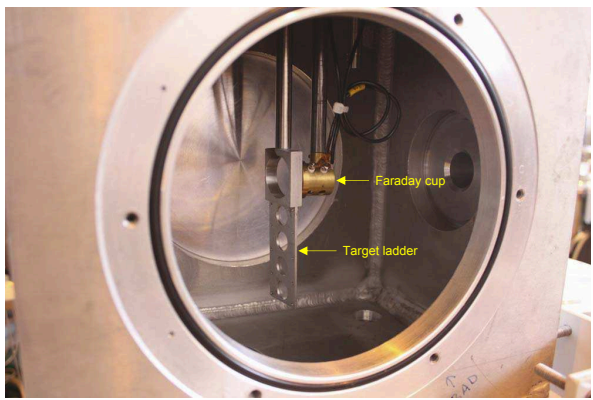}
\caption{\label{scatchamber}Interior of the target chamber used in the experiment. The target ladder and the Faraday cup are indicated.}
\end{figure}

The present experiment utilized the 20$^{\circ}$ beam line of the REX-ISOLDE facility at CERN. The 2.82 MeV/u $^{129}$Xe and $^{78}$Kr ion beams were produced in the ISOLDE GPS target and accelerated with the REX - ISOLDE linear accelerator. The $^{78}$Kr beam contained an admixture of 2.82 MeV/u $^{49}$Ti ions which were detected with PET and which provided an additional data point for calibration. 
The target chamber for carrying out the implantations is shown in Fig.~\ref{scatchamber}.

The target ladder seen in this figure has four circular apertures of 15 mm diameter which allows for mounting and exposure of the samples and a larger opening of diameter 30 mm behind which a Faraday cup could be mounted for monitoring beam current. The PET films were cut into rectangular strips (8.0 cm $\times$ 3.5 cm) so that it covered all four apertures and fixed onto the downstream side of the target ladder with screws. This ensured that the exposed areas were well defined. The target ladder and the Faraday cup were mounted on actuators. This allowed for the monitoring of the beam current as well as upto four different exposures of PET samples without venting the target chamber, by moving the actuator up or down. The actuator to which the target ladder was mounted could also be rotated. This enabled us to vary the angle of incidence of the ions on PET films. To ensure that the number of the impinging ions lie in the range mentioned before, beam currents of 1.5 pA for $^{78}$Kr and 3.1 pA for $^{129}$Xe beams were first measured with the Faraday cup, because for too low beam currents direct measurements are not possible. The beam intensities were then attenuated by a factor of 0.005 by employing the REXTRAP beam gate. This made sure that for exposure duration of 30 sec, we got $\sim 5\times 10^4$ particles/cm$^2$.

In all 20 different exposures were carried out. For half of the exposures the angle of incidence was kept at 30$^{\circ}$ with respect to the beam. This was done to make sure that the conical profile of the tracks are clearly visible after etching, thereby enabling a more accurate determination of track parameters. Typical pressure inside the target chamber during the experiment was kept at $10^{-5}$ mbar.

\begin{figure}[t]
\includegraphics[width = 1.0\hsize,clip]{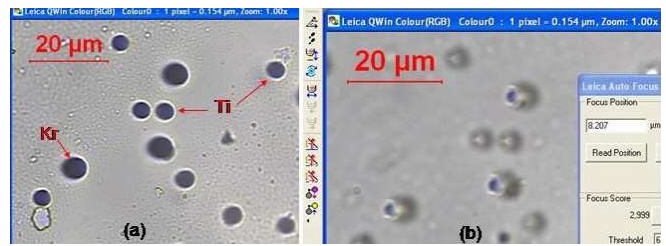}
\caption{\label{krti}The different track dimensions for $^{78}$Kr and $^{49}$Ti can be seen with the microscope focussed (a) on the surface and (b) at a depth of 8.2 $\mu$m.}
\end{figure}

The exposed PET samples were then etched in 6.25 N NaOH solution at 55.0$ \pm $0.5$^{\circ}$C which was found~\cite{maulik,nimb} to be the ideal combination for etching PET detectors. The etched samples were then studied under $\times$100 dry objective of a Leica digital microscope which was interfaced with a computer preloaded with an image analysis software. Fig.~\ref{krti} shows track images due to $^{78}$Kr as well as $^{49}$Ti on PET after 2 hr etching. Different track diameters corresponding to the two ions are evident from Fig~2(a) with the microscope focussed on the detector surface. Fig~2(b) shows that the end of track is reached for $^{49}$Ti ions but not for $^{78}$Kr with the microscope focussed at a depth of 8.2 $\mu$m. This implies different track lengths for the two ions.

\vspace{0.3 cm}
\centerline{\bf{RESULTS AND DISCUSSION}}
\vspace{0.3cm}

\begin{figure}[t]
\includegraphics[width =1.0\hsize,clip]{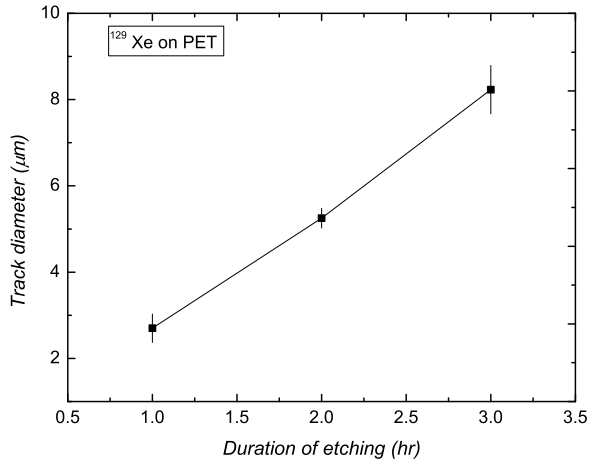}
\caption{\label{xediavar}The variation of track diameter for $^{129}$Xe tracks in PET for different durations of etching. The line is only to guide the eye.}
\end{figure}

\begin{figure}[t]
\includegraphics[width = 1.0\hsize,clip]{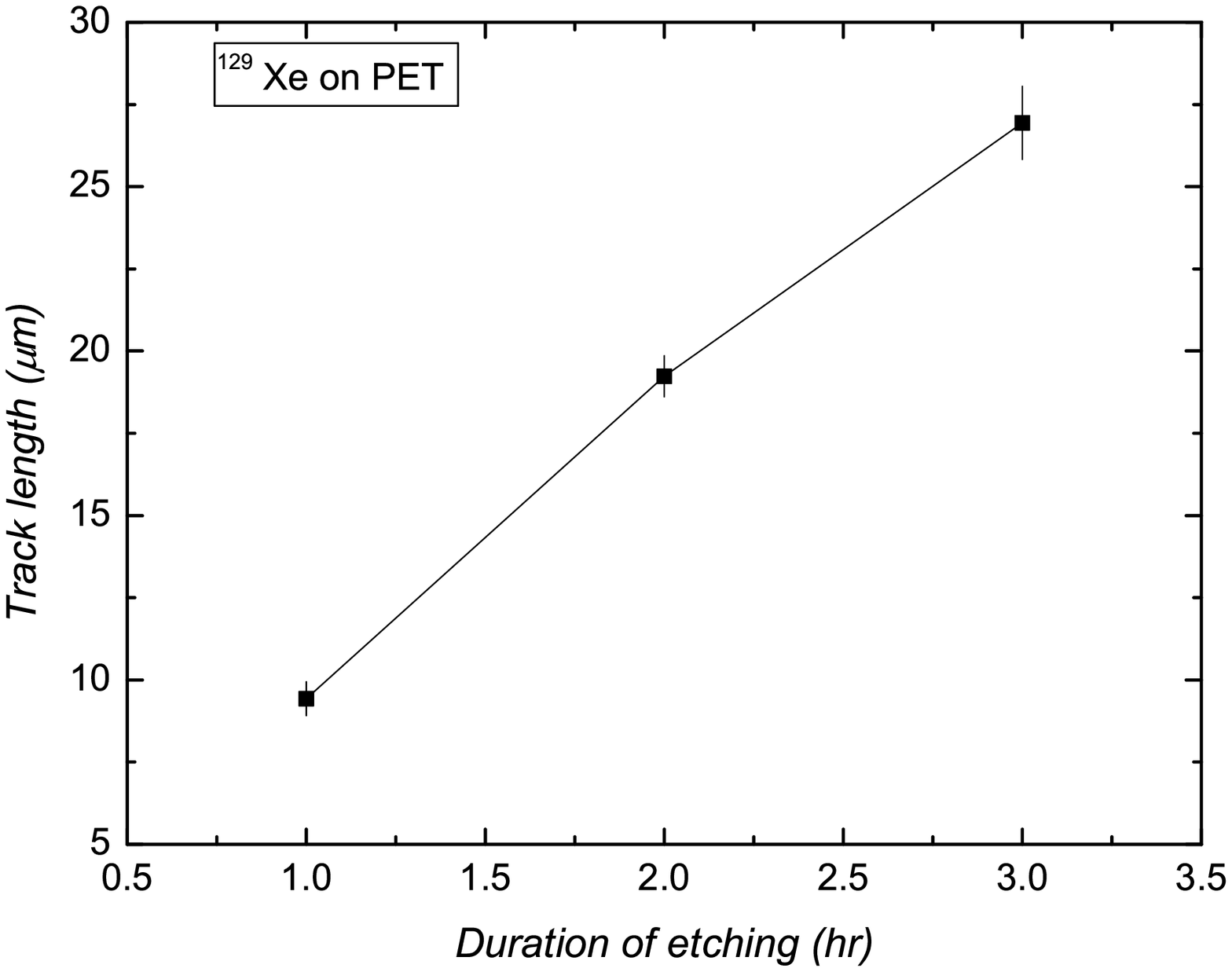}
\caption{\label{xetralenvar} The variation of track length for $^{129}$Xe tracks on PET for different durations of etching. The line is only to guide the eye.}
\end{figure}

\begin{figure}[t]
\includegraphics[width = 1.0\hsize,clip]{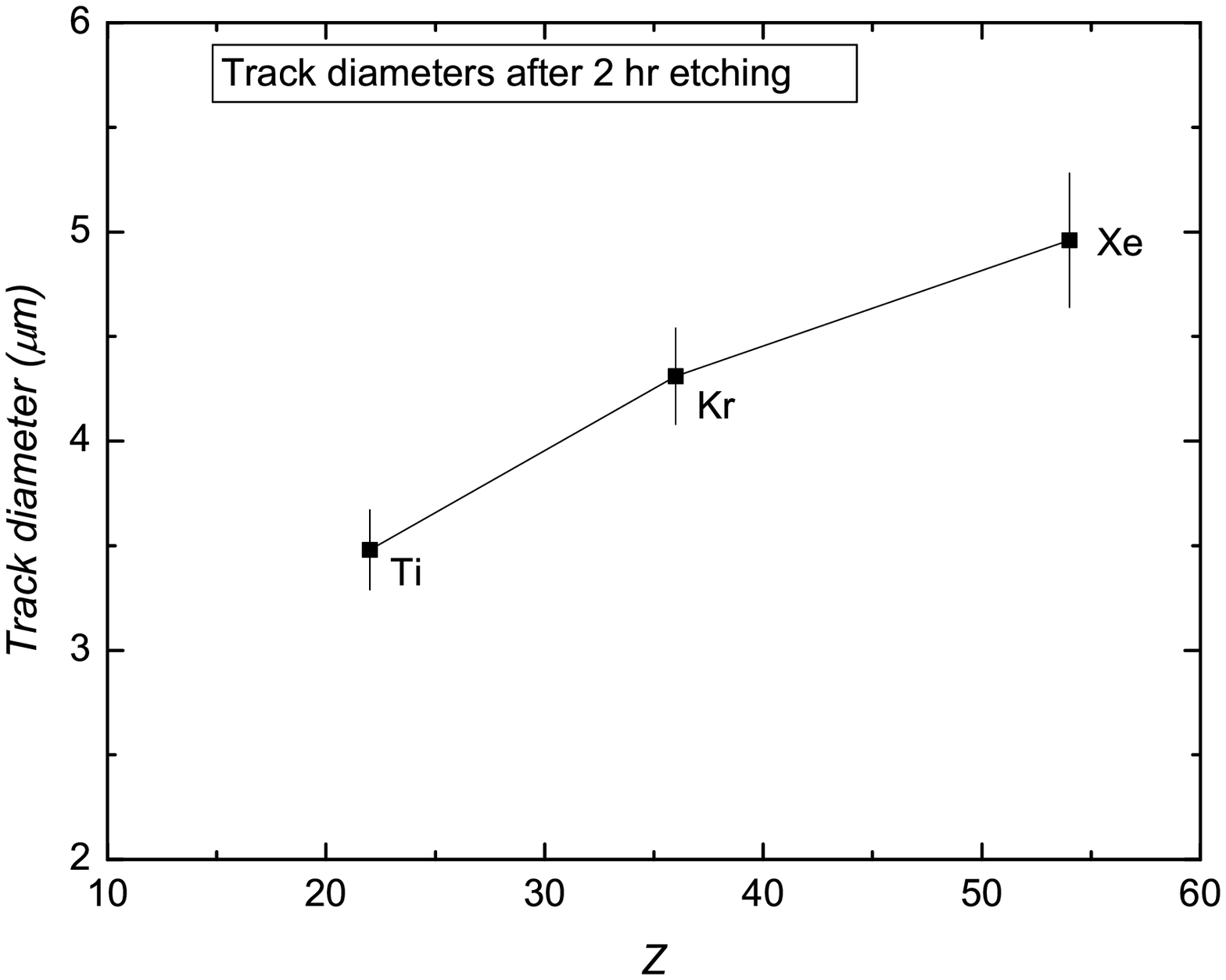}
\caption{\label{tradiacomp}Figure compares the diameters of $^{129}$Xe, $^{78}$Kr and $^{49}$Ti tracks on PET after 2 hr etching. The line is only to guide the eye.}
\end{figure}

\begin{figure}[t]
\includegraphics[width = 1.0\hsize,clip]{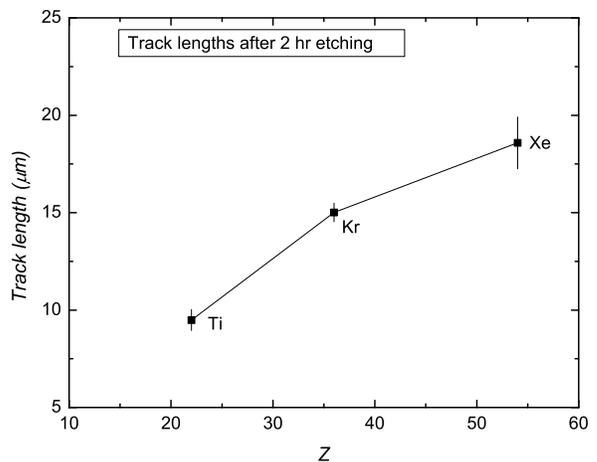}
\caption{\label{tralencomp}Figure compares the track lengths of $^{129}$Xe, $^{78}$Kr and $^{49}$Ti tracks on PET after 2 hr etching. The line is only to guide the eye.}
\end{figure}

Figs.~\ref{xediavar} and \ref{xetralenvar} show the variation of track diameter and length with duration of etching for tracks due to $^{129}$Xe ions of energies 2.82 MeV/u. The monotonic rise with etching time of the track diameters and lengths implies a consistency in the rate of etching for PET. Figs.~\ref{tradiacomp} and \ref{tralencomp} compare the track diameters and track lengths for the three different ions, all with energies of 2.82 MeV/u, for a fixed etching duration. The same energy per nucleon implies the same values for $\beta$, thus the differences seen in track diameters and lengths purely reflect the different Z values for the three ions. This can provide a method of identifying incident particles when they have similar values for energy per nucleon. 

\begin{figure}[t]
\includegraphics[width = 1.0\hsize,clip]{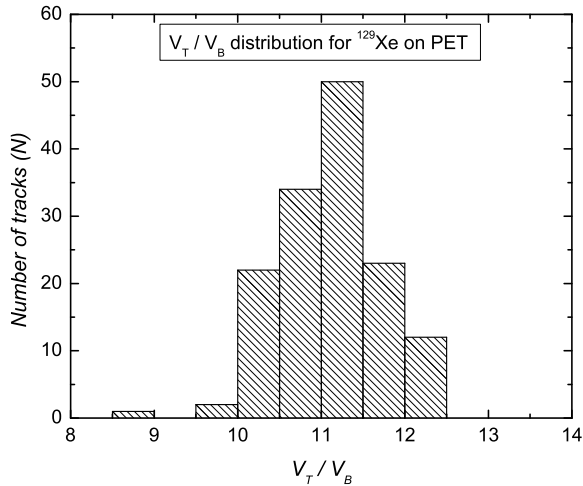}
\caption{\label{xevtvbhist}Histogram showing the distribution of $V_T/V_B$ values corresponding to $^{129}$Xe tracks on a PET sample.}
\end{figure}

Fig.~\ref{xevtvbhist} shows the distribution of $V_T/V_B$ values for $^{129}$Xe obtained from track parameter measurements on a particular PET sample. The closely bunched values of $V_T/V_B$ for $^{129}$Xe show the consistency of the charge response of PET. 

\begin{figure}[t]
\includegraphics[width = 1.0\hsize,clip]{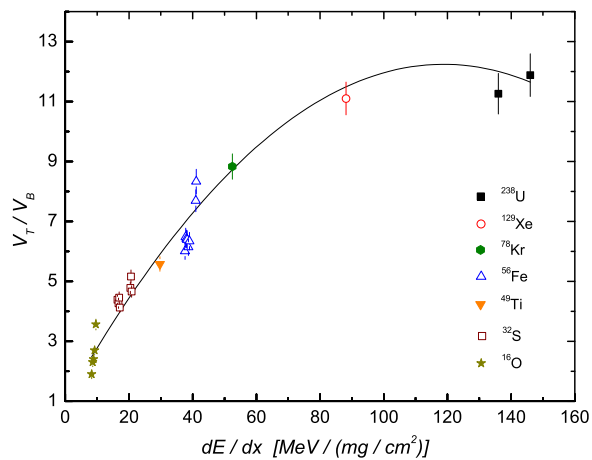}
\caption{\label{calibration}Calibration curve for PET.\\}
\end{figure}

By combining the $V_T/V_B$ values for $^{129}$Xe, $^{78}$Kr, $^{49}$Ti as well as those for $^{16}$O, $^{32}$S, $^{56}$Fe, $^{238}$U ions obtained earlier ~\cite{maulik, nimb}, we can get a calibration curve for PET as shown in Fig.~\ref{calibration}. The specific energy losses $dE/dx$ of the incident ions at different energies were obtained using the Monte Carlo code SRIM ~\cite{ZI03}. 

The nature of the calibration curve can be explained thus. High values of $dE/dx$ leads to highly damaged polymer matrix along the latent track. But the etching process is limited by the transportation of etching solution and etching products in the narrow track channel~\cite{West94}\cite{Enge95}. So for higher $dE/dx$ values, large quantities of etching products cannot be removed from the track channel fast enough. So $V_T$ starts to slow down, will reach a maximum and may then fall for any further increase in $dE/dx$. Consequentely the ratio $V_T/V_B$ is also expected to show similar behavior. The fitted line is given by $y= a + bx + cx^2$,  where $a = 0.96\pm0.08, b = 0.19\pm0.01$ and $c = (0.80\pm0.05)\times10^{-3}$ where the parameter errors represent 95\% confidence level. With this calibration curve one can find $dE/dx$ of any charged particle impinging on PET. Along with the measured range inside the PET detector for a stopping charged particle, one can identify any charged particle with its charge, mass and energy. \\

\section{Conclusion}
Our work has clearly demonstrated that PET can be effectively used as a charged particle detector with a high detection threshold. This could be particularly useful in cases where there is a requirement for detecting high Z particles against a low Z background such as rare event search in presence of dominant protons in cosmic rays. Another advantage for PET is its significantly low cost compared to other commercially available SSNTDs. Cost considerations become particularly important if one intends to set up large area passive detector arrays. We plan to setup a large array of PET at Sandakphu (alt. 3600~m) in Eastern Himalayas to look for strangelets in cosmic rays. 

\begin{acknowledgments}
The authors sincerely thank Dr. Y. Blumenfeld and Dr. A. Herlert of ISOLDE, CERN for their help in getting the PET films irradiated. The authors also thank Mr. Sujit K. Basu for technical assistance. This work was performed under the aegis of the IRHPA (Intensification of Research in High Priority Areas) Project (IR/S2/PF-01/2003) of the Science and Engineering Research Council (SERC), DST, Government of India, New Delhi. One of the authors, JP, has been supported by a Marie Curie Intra-European Fellowship of the European Community's 7th Framework Programme under contract number PIEFGA-2008-219175.
\end{acknowledgments}

\end{document}